\documentclass{appolb}
\usepackage{graphicx}
\begin{document}
\title{Measurement of global spin alignment of vector mesons at RHIC%
\thanks{Presented at the 29$^{\mathrm{th}}$ International Conference on
  Ultra-relativistic Nucleus-Nucleus Collisions (Quark Matter 2022),
  Krakow, Poland. SS is supported by the Strategic Priority Research Program of Chinese Academy of Sciences. }%
}
\author{Subhash Singha (For STAR Collaboration)
\address{Institute of Modern Physics Chinese Academy of Sciences,
  Lanzhou, China 73000\\
(subhash@impcas.ac.cn)}
%\\[3mm]
%{Third Author % of different affiliation
%\address{affiliation}
%}
%\\[3mm]
%the Name(s) of other Author(s)
%\address{affiliation}
}
\maketitle

%\begin{linenumbers}

\begin{abstract}
% length not exceeding 200 words
%Here comes the abstract

We report the measurements of spin alignment ($\rho_{00}$) for $K^{*0}$,
$\overline{K^{*0}}$, $K^{*+}$, and $K^{*-}$ vector mesons in RHIC
isobar collisions (Zr+Zr and Ru+Ru) at $\sqrt{s_{\mathrm {NN}}}$ = 200
GeV. We observe the first non-zero
spin alignment for $K^{*\pm}$ in heavy-ion collisions. 
%At mid-central collisions, the $\rho_{00}$ of both charged and
%neutral $K^{*}$ species are larger than $\frac{1}{3}$. 
The  $K^{*\pm}$ $\rho_{00}$ is about 3.9$\sigma$ larger than that of
$K^{*0}$. The observed difference and the ordering between $K^{*\pm}$ and $K^{*0}$ are
surprising, and require further inputs from theory. When
comparing between the isobar and Au+Au collisions, no 
significant system size dependence in $K^{*0}$ $\rho_{00}$ is observed within uncertainties.

\end{abstract}
 
% for parallel talks 6 pages in total
\section{Introduction}
In the initial stage of non-central heavy-ion collisions (HIC), a large
orbital angular momentum (OAM) is imparted into the system. The magnitude of such OAM
can be $ \sim bA \sqrt{s_{\mathrm {NN}}} \sim 10^{4} \hbar$, where $b$ is the
impact parameter and $A$ is the mass number of the collision species~\cite{becattini}. A
part of OAM transferred to the Quark Gluon Plasma (QGP) medium can  polarize quarks and anti-quarks due to
``spin-orbit'' interaction and hence induce
a non-vanishing polarization for hadrons with non-zero spin~\cite{liang0}. The
incoming charged spectators in HIC can also induce a large but
short lived magnetic field ($eB \sim 10^{18} 
\rm$ Gauss)~\cite{kharzeev}. Such a strong $B$-field can also polarize
both quarks and anti-quarks due to its coupling with the intrinsic magnetic moment. The
 measurement of spin polarization can not only offer insights into the
 initial orbital angular momentum interactions and magnetic field, but
 also serve as an experimental probe to understand the response of QGP medium
 under these extreme initial conditions. 
%It also provide a unique opportunity to probe the spin degrees of freedom of the QGP. 
The measurement of significant non-zero polarization of $\Lambda$ hyperons by STAR
 collaboration offered first experimental evidence of the presence of 
 vorticity of the QGP medium induced by the initial angular
 momentum, while a hint of difference between $\Lambda$  and
 $\overline{\Lambda}$ spin polarization at RHIC presents an opportunity to probe
 the initial $B$-field~\cite{star_lambda_nature}. 
%The vector meson's spin alignment can also be used to probe both the vorticity dynamics and initial $B$ field. 

The spin alignment is quantified by $00^{\mathrm{th}}$ element of
 the spin density matrix, $\rho_{00}$, and can be measured from the angular
 distribution of the decay daughter of the vector meson~\cite{schiling}:
\begin{equation}
  \frac{d\rm{N}}{d\rm{cos}\theta^{*}} \propto \bigg(   (1-\rho_{00}) +
  (3\rho_{00}-1) \rm{cos}^{2} \theta^{*} \bigg),
\label{eqn1}
\end{equation}
where $\theta^{*}$ is the angle between the polarization axis and
momentum direction of the daughter particle in
the rest frame of its parent. For global spin alignment, the
polarization axis is chosen as the direction perpendicular to the
reaction plane, which can be correlated with both the OAM and the $B$-field. The
value of $\rho_{00}$ is expected to be $\frac{1}{3}$ in absense of spin
alignment, while a deviation of $\rho_{00}$ from $\frac{1}{3}$ indicates a net spin alignment. 

%The theory of $\rho_{00}$ can be complicated and several calculations
%indicated that there are different physics mechanisms can alter the
%deviation of $\rho_{00}$ differently.
At present, the available physics mechanisms that can cause
spin alignment are the following:
$(i)$ the  polarized quarks induced by vorticity can
hadronize via coalescence mechanism. It can make $\rho_{00}$ smaller than $\frac{1}{3}$~\cite{liang0, yang};
$(ii)$ the $\rho_{00}$ induced by the $B$-field can be either larger or
smaller than $\frac{1}{3}$. The expected deviation due to vorticity and
$B$-field is $\rho_{00} - \frac{1}{3} \sim 10^{-5}$~\cite{yang};
$(iii)$ the electric field can give a positive contribution with
$\rho_{00} - \frac{1}{3} \sim 10^{-4}$~\cite{yang};
$(iv)$ the fragmentation of polarized quarks can make either positive or
negative contribution with $\rho_{00} - \frac{1}{3} \sim 10^{-5}$~\cite{liang0};
$(v)$ local spin alignment, helicity polarization, and 
turbulent color field can also make $\rho_{00}$ smaller than $\frac{1}{3}$~\cite{xia_guo_muller};
$(vi)$ a fluctuating strong force field of vector meson can cause the
$\rho_{00}$ to be larger than $\frac{1}{3}$ with a deviation $\sim
0.1$, which is an order of magnitude larger compared to more conventional
mechanisms~\cite{sheng}. 
The study of $\rho_{00}$ of various vector meson species can thus
elucidate our understanding of different mechanisms causing spin
alignment. Furthermore, the neutral and charged
vector mesons ($K^{*0}(d\bar{s})$ and $K^{*+}(u\bar{s})$) have similar
mass, but the magnetic moments of their constituent quarks differ by
about a factor of five ($\mu_{d} \sim -0.97 \mu_{\mathrm{N}}$, $\mu_{u} \sim 1.85
\mu_{\mathrm{N}}$). Hence, the magnetic field driven
contribution to the $\rho_{00}$ of neutral and charged $K^{*}$ is
expected to be different.

The recent measurements of $\rho_{00}$ of $\phi$ and $K^{*0}$ vector
mesons from the 1$^{st}$ phase of RHIC Beam
Energy Scan (BES-I) Au+Au collisions revealed a surprising
pattern~\cite{star_bes_rho00}. While the $K^{*0}$ $\rho_{00}$ is
largely consistent with $\frac{1}{3}$, the $\phi$ mesons show a large positive
deviation ($\rho_{00} > \frac{1}{3}$) with 8.4$\sigma$ significance when
$\rho_{00}$ is integrated within the range $\sqrt{s_{\mathrm {NN}}} =
11.5 - 62.4$ GeV for $1.2 < p_{\mathrm{T}} < 5.4$ GeV/$c$ in 20-60\%
Au+Au collisions. Such a large positive deviation at mid-central
collisions pose challenges to more conventional physics
mechanisms, while the polarization induced from a fluctuating
$\phi$-meson vector field can accommodate the large positive
signal~\cite{sheng}. Moreover, the $p_{\mathrm{T}}$ and centrality differential
measurements of $\phi$ and $K^{*0}$ $\rho_{00}$ in BES-I energy range also show
non-trivial patterns~\cite{star_bes_rho00}. 
% These differential measurements need confrontation

\section{Analysis method}
This proceedings report the first $\rho_{00}$ measurements of charged $K^{*\pm}$ along
with neutral $K^{*0}$ ($\overline{K^{*0}}$) vector mesons in RHIC isobar collisions of
$_{44}^{96}$Ru+$_{44}^{96}$Ru and $_{40}^{96}$Zr+$_{40}^{96}$Zr species
at $\sqrt{s_{\mathrm {NN}}}$ = 200 GeV~\cite{blind}. The $K^{*0}(\overline{K^{*0}})$ and
$K^{*+}(K^{*-})$ are reconstructed via $K^{*0}(\overline{K^{*0}}) \rightarrow \pi^{-}+K^{+}
(\pi^{+}+K^{-})$ and $K^{*+}(K^{*-}) \rightarrow \pi^{+}+K_{\mathrm{S}}^{0}
(\pi^{-}+K_{\mathrm{S}}^{0})$ respectively. 
The minimum-bias (MB) events are collected
via a coincidence between the Vertex Position Detectors (VPD) located at $4.4 <
|\eta| < 4.9$. For analysis, the vertex position along the beam
($V_{\mathrm{z,TPC}}$) and
radial direction ($V_{r}$) are required to be
within $-35 < V_{\mathrm{z,TPC}} < 25$ cm and $V_{r} < $~5 cm respectively with
a coordinate system at the center of Time Projection Chamber (TPC).  
%A good quality events are ensured by selecting $ |V_{z,TPC} -
%V_{z,VPD}| <$ 5 cm, where $V_{z,VPD}$ is the vertex position along
%beam direction reconstructed using the VPDs. The events with pile-up (about 0.5\%) are removed by
%excluding outliers in the correlation between the number of TPC
%tracks and the number of those tracks matched with a hit in the Time
%Of Flight (TOF) detector. 
We analyzed about 1.8 and 2.0 billion good MB events for Ru+Ru
and Zr+Zr collisions, respectively.  The charged particle tracking is performed using the TPC. The collision centrality is determined from the number of charged particles within $|\eta| <$ 0.5,
and using a Monte Carlo Glauber simulation~\cite{centrality_glauber}. The second order event
plane ($\Psi_{2,\mathrm{TPC}}$) is reconstructed using the tracks inside
TPC~\cite{flow}. In isobar collisions, the typical $\Psi_{2,\mathrm{TPC}}$ resolution achieved in
mid-central collisions is $R_{2,\mathrm{TPC}} \sim $ 64\%. The decay daughters
of $K^{*}$ are identified using the specific ionization energy loss in
TPC gas volume and the velocity of particles measured by the TOF detector. The 
$K_{\mathrm{S}}^{0}$ mesons are selected via a weak decay topology. For charged
$K^{*\pm}$ reconstruction, only the $K_{\mathrm{S}}^{0}$ candidates within $0.48 <
M(\pi^{+} \pi^{-})< 0.51$ GeV/$c$$^{2}$ are considered. The combinatorial
background is estimated from a track rotation technique, in which one
of the daughter track is rotated by 180$^\circ$ to break the
correlation among the pairs originating from same parent particle. Then,
the invariant mass signal is obtained by subtracting the combinatorial background. The $K^{*}$ signal is
fitted with a Breit-Wigner distribution and a second-order polynomial
function to take care of residual background. 

\begin{figure}[htb]
%\centerline{%
%}
\includegraphics[scale=0.33]{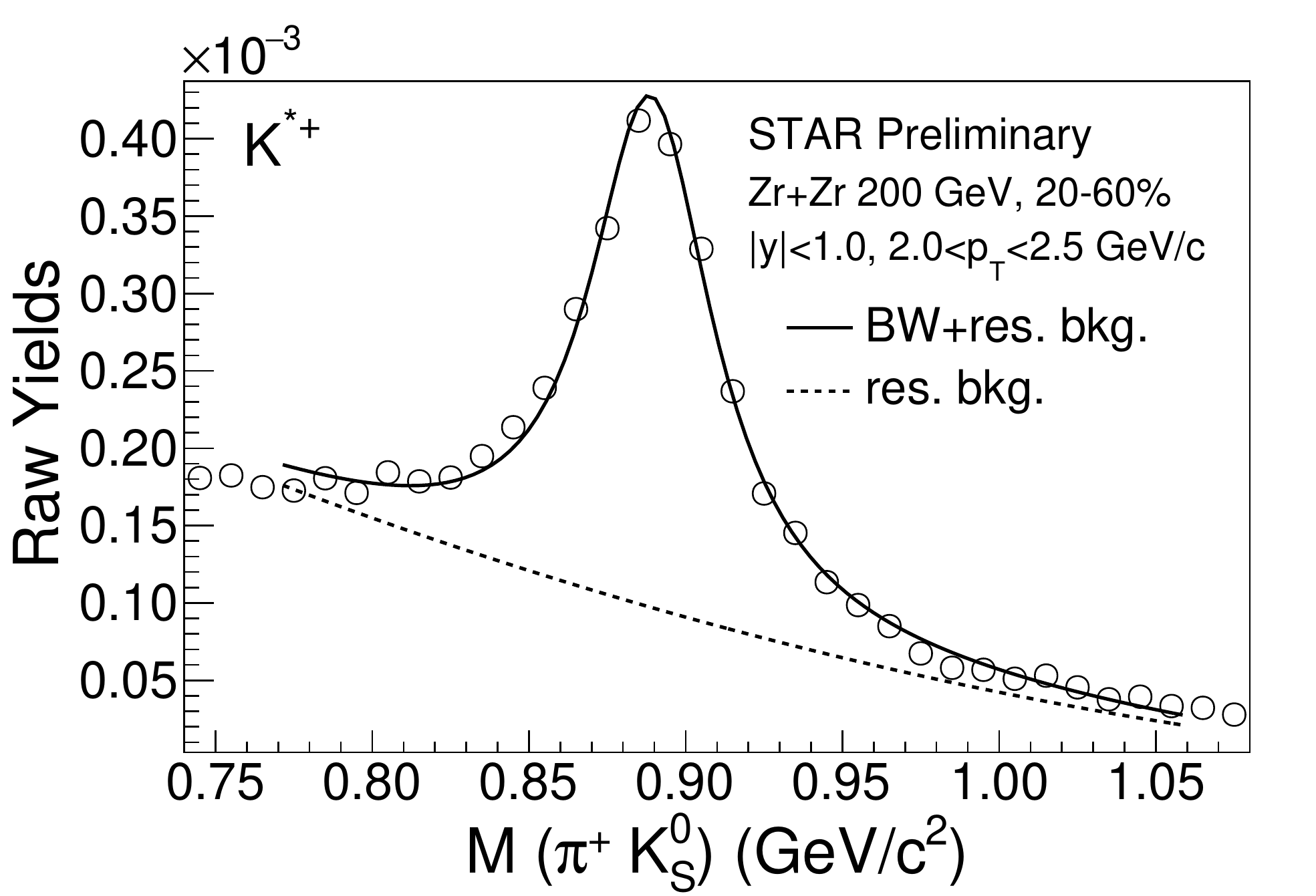}
\hspace{-0.2cm}
\includegraphics[scale=0.33]{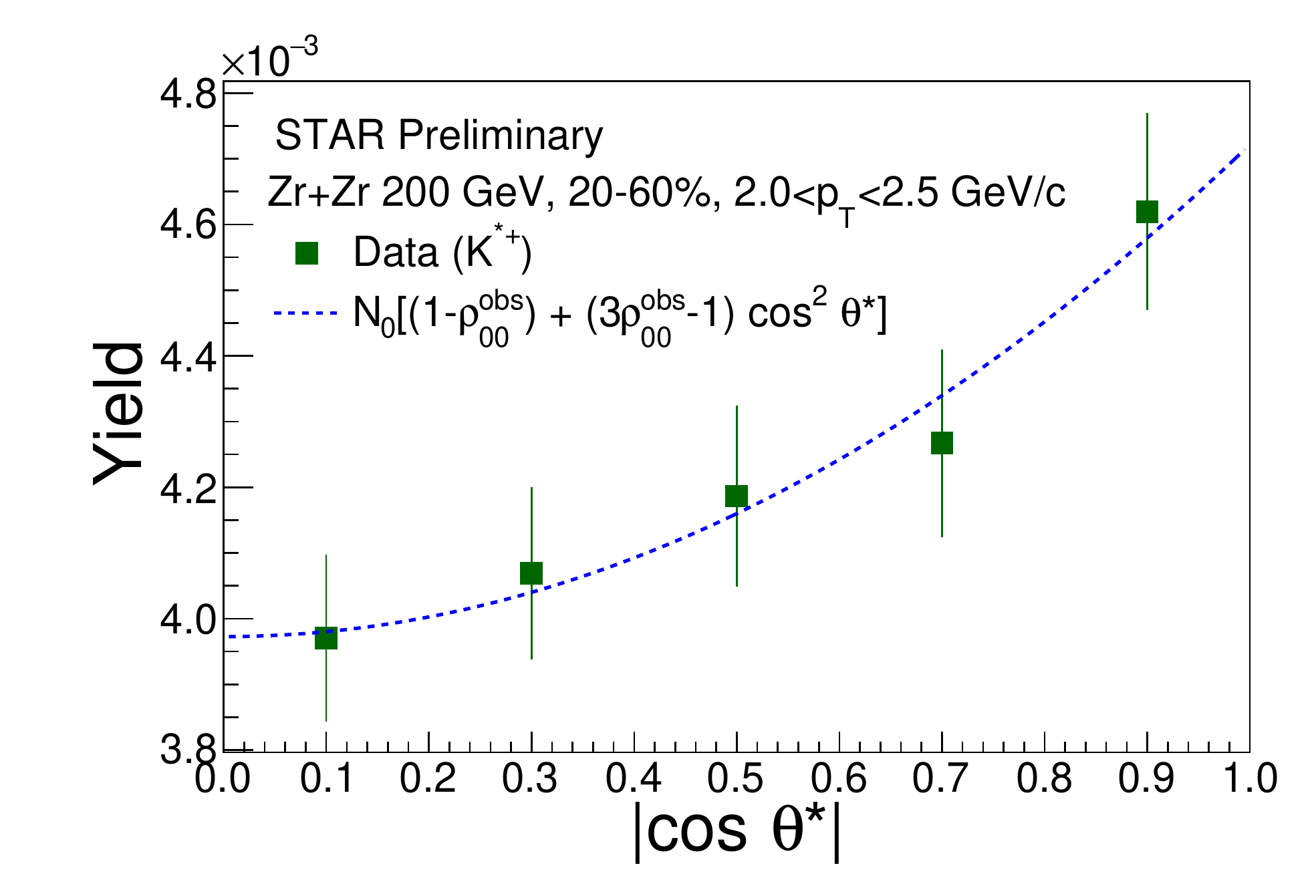}
\vspace{-0.5cm}
\caption{Left:  $K^{*+} (\rightarrow \pi^{+} + K_{\mathrm{S}}^{0})$ invariant mass distribution for $2.0 <
p_{\mathrm{T}} < 2.5$ GeV/$c$ in 20-60\%  Zr+Zr collisions at $\sqrt{s_{\mathrm {NN}}}$ = 200
GeV. Right: efficiency and
acceptance corrected $K^{*+}$ yield as a function
of $|\rm{cos}\;\theta^{*}|$ in 200 GeV Zr+Zr collisions.}
\label{Fig1_signal_thetafit}
\end{figure}

The left panel in Fig.~\ref{Fig1_signal_thetafit} presents the $K^{*+}$ signal for $2.0 <
p_{\mathrm{T}} < 2.5$ GeV/$c$ in 20-60\%  Zr+Zr collisions at $\sqrt{s_{\mathrm {NN}}}$ = 200 GeV.
The yield is estimated by integrating residual background subtracted
signal within the range: $m_{0} \pm 3\Gamma$, where
$m_{0}$ and $\Gamma$ are the invariant mass peak position and width of
$K^{*}$. The yield is obtained in five $|\rm{cos}\;\theta^{*}|$ bins where $\theta^{*}$ is the angle between
$\Psi_{2,\mathrm{TPC}}$ and momentum of daughter kaon (pion) in parent
$K^{*0}$ ($K^{*\pm}$) rest frame. 
The detector acceptance and efficiency correction factors are obtained
using a STAR detector simulation in GEANT3. The right panel in
Fig.~\ref{Fig1_signal_thetafit} presents efficiency and
acceptance corrected $K^{*+}$ yield as a function
of $|\rm{cos} \; \theta^{*}|$ for $2.0 < p_{\mathrm{T}} < 2.5$ GeV/$c$ in 20-60\%  Zr+Zr
collisions. The yield versus $|\rm{cos} \; \theta^{*}|$ distribution is then fitted with
Eq.\ref{eqn1} and the extracted $\rho_{00}$ (called
$\rho_{00}^{\mathrm{obs}}$) is corrected for event plane resolution using:
$\rho_{00} = \frac{1}{3} + \frac{4}{1+3R_{2,\mathrm{TPC}} }(\rho_{00}^{\mathrm{obs}} -
\frac{1}{3})$~\cite{aihong}.
%\begin{equation}
%\rho_{00} = $\frac{1}{3}$ + \frac{4}{1+3R}(\rho_{00}^{obs} - $\frac{1}{3}$)
%\label{eqn2}
%\end{equation}

\section{Results}
\begin{figure}[htb]
%\centerline{%
%}
\includegraphics[scale=0.37]{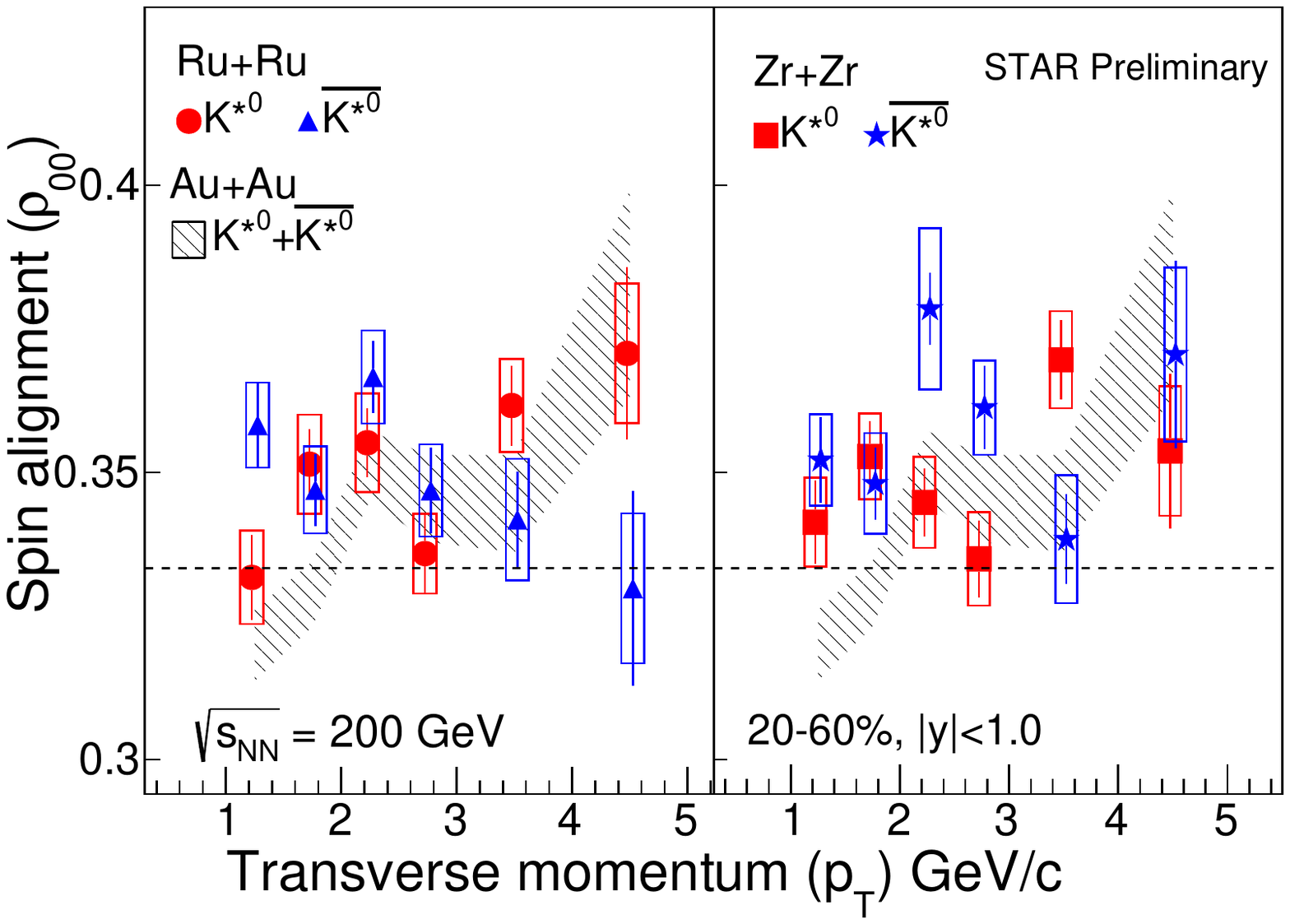}
\hspace{-1.2cm}
\includegraphics[scale=0.37]{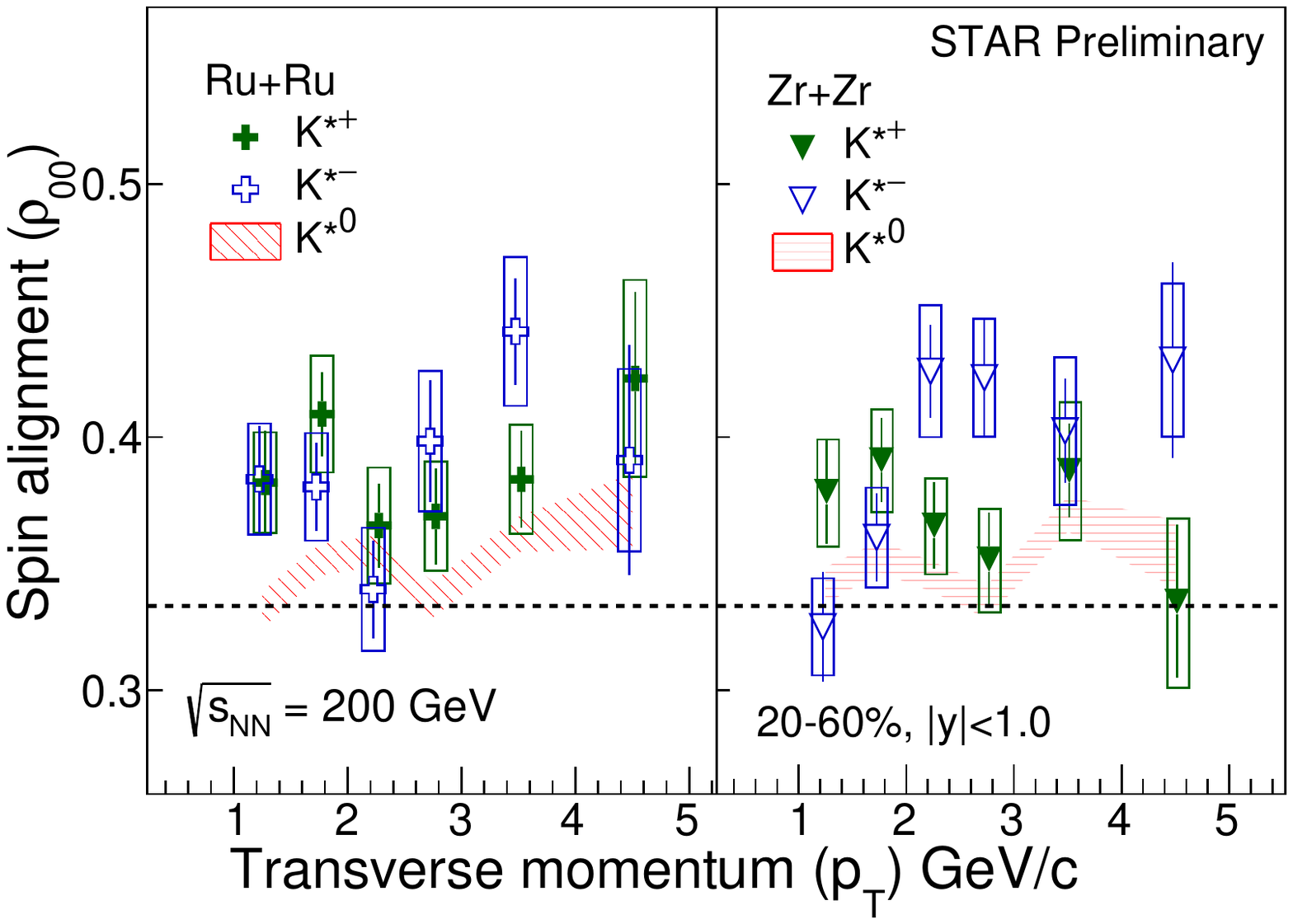}
\vspace{-1.0cm}
\caption{Left: $\rho_{00}(p_{\mathrm{T}})$ for $K^{*0}$ and $\overline{K^{*0}}$
  in isobar collisions at $\sqrt{s_{\mathrm {NN}}}$  = 200 GeV. Results are
  compared with that from 200 GeV Au+Au
  collisions~\cite{star_bes_rho00}. Right: Comparison of
  $\rho_{00}(p_{\mathrm{T}})$ between $K^{*\pm}$ and $K^{*0}$ in 200 GeV isobar
  collisions.}
\label{Fig2_rho00_pT}
\end{figure}

\begin{figure}[htb]
%\centerline{%
%}
\includegraphics[scale=0.37]{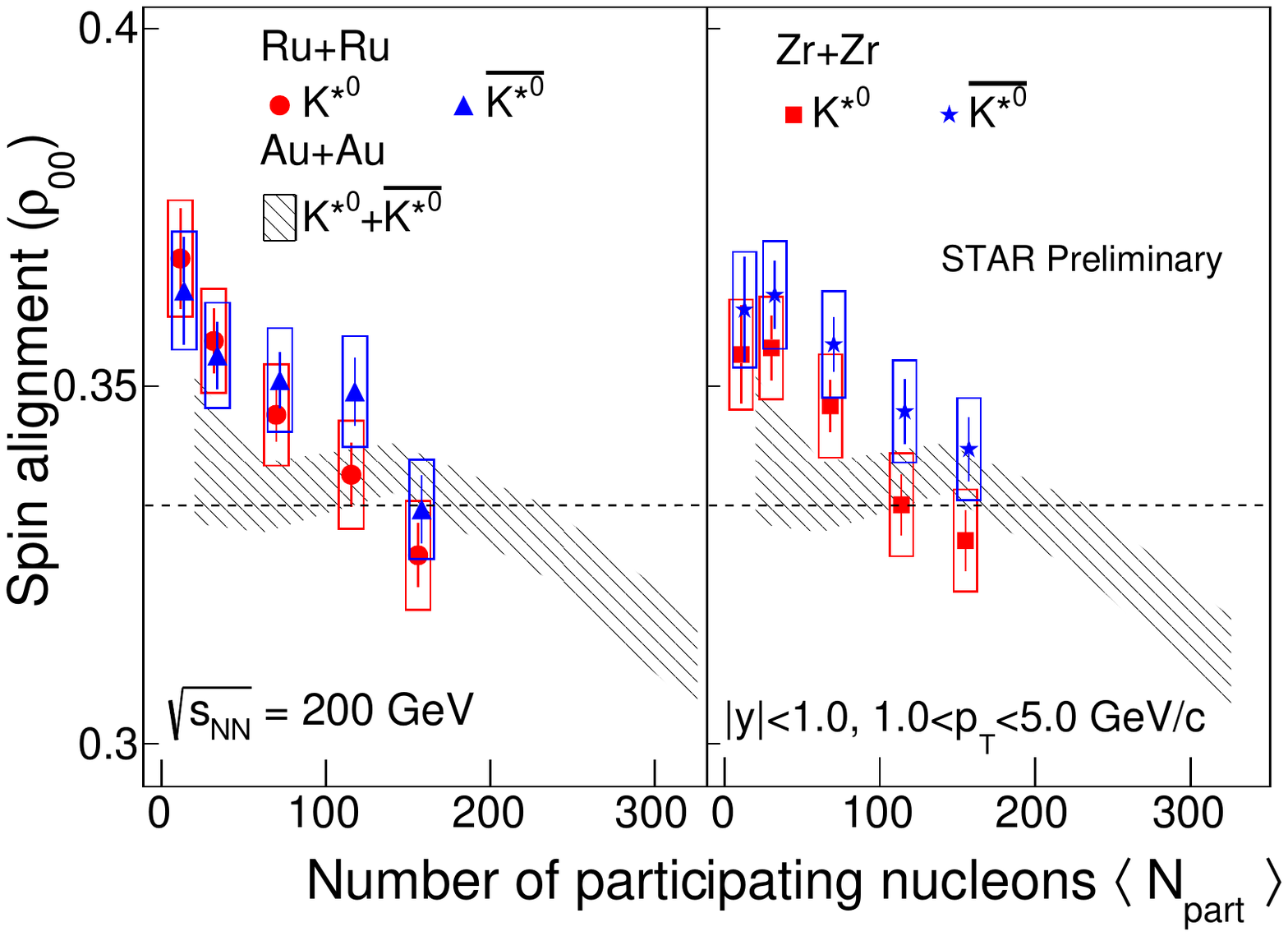}
\hspace{-1.0cm}
\includegraphics[scale=0.37]{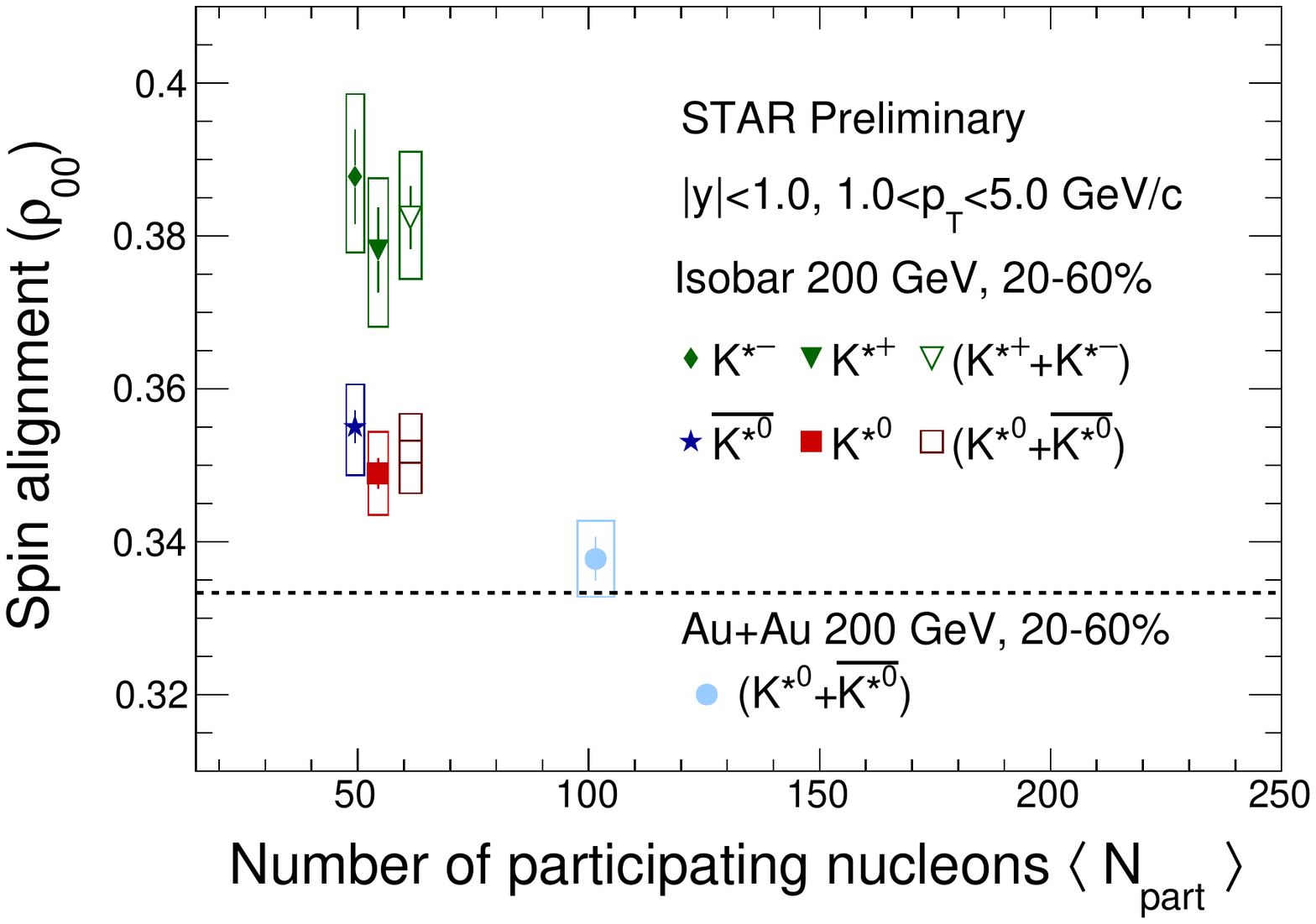}
\vspace{-0.7cm}
\caption{Left: $\rho_{00}(\langle N_{\mathrm{part}} \rangle)$ for $K^{*0}$ and $\overline{K^{*0}}$
  in isobar collisions at $\sqrt{s_{\mathrm {NN}}}$ = 200 GeV. Right: $p_{\mathrm{T}}$
  integrated $\rho_{00}$ for $K^{*0}$, $\overline{K^{*0}}$, $K^{*+}$ and $K^{*-}$
  in 20-60\% 200 GeV isobar collisions. Results are compared with
  $K^{*0}$ in 200 GeV Au+Au
  collisions~\cite{star_bes_rho00}.}
\label{Fig3_rho00_cent}
\end{figure}

The left panel of Fig.~\ref{Fig2_rho00_pT} presents the $p_{\mathrm{T}}$
dependence of $\rho_{00}$ for $K^{*0}$ and $\overline{K^{*0}}$ at
mid-rapidity ($|y|<1.0$) in 20-60\% central Ru+Ru and Zr+Zr collisions
at $\sqrt{s_{\mathrm {NN}}}$ = 200 GeV. The $\rho_{00}$ between the
particle and anti-particle species are consistent within errors.
These results are compared with that from 200 GeV
Au+Au collisions~\cite{star_bes_rho00}. The $\rho_{00}$ between isobar
and Au+Au collisions are consistent within uncertainties across the measured
$p_{\mathrm{T}}$ region in mid-central collisions. The right panel of Fig.~\ref{Fig2_rho00_pT} shows a
  comparison of $\rho_{00}(p_{\mathrm{T}})$ among neutral and charged $K^{*}$
  species in isobar collisions. The $\rho_{00}$ for charged $K^{*\pm}$
  are systematically larger than the neutral $K^{*0}$ across the
  measured $p_{\mathrm{T}}$ region. The left panel of Fig.~\ref{Fig3_rho00_cent}
  presents the $\rho_{00}$ as a function of average number of
  participants ($\langle N_{\mathrm{part}} \rangle$) for $K^{*0}$ and
  $\overline{K^{*0}}$ for $1.0 < p_{\mathrm{T}} < 5.0$ GeV/$c$ in 200 GeV Ru+Ru and Zr+Zr
  collisions. These results are compared with that from 200 GeV Au+Au
  collisions~\cite{star_bes_rho00}.  The $K^{*0}$ $\rho_{00}$ is
  larger than $\frac{1}{3}$ at smaller $\langle N_{\mathrm{part}} \rangle$. It
  is smaller than $\frac{1}{3}$ at large $\langle N_{\mathrm{part}} \rangle$, which can
  have contributions from the local spin
  alignment~\cite{xia_guo_muller}. At a similar $\langle N_{\mathrm{part}}
  \rangle$,  the $\rho_{00}$ between small system isobar and large
  system Au+Au are comparable within uncertainties.

The right panel of Fig.~\ref{Fig3_rho00_cent} summarizes the
$p_{\mathrm{T}}$-integrated $\rho_{00}$ for $K^{*0}$, $\overline{K^{*0}}$, $K^{*+}$ and $K^{*-}$
  in 20-60\% isobar collisions. These results are compared with
  ($K^{*0}$+$\overline{K^{*0}}$) $\rho_{00}$ from Au+Au
  collisions~\cite{star_bes_rho00}. This is the first
  observation of $K^{*\pm}$ $\rho_{00}$ to be larger than $\frac{1}{3}$ in
  heavy-ion collisions. Moreover, the $p_{\mathrm{T}}$-integrated $\rho_{00}$ reveals a clear ordering
between neutral and charged $K^{*}$ species in isobar collisions, with
the charged species about 3.9$\sigma$ larger than the neutral
ones. Due to the interaction between the $B$-field and the magnetic
moment of the constituent quarks, one naively expects the $K^{*0}$
$\rho_{00}$ to be larger than that of $K^{*\pm}$~\cite{yang}. But the
observed ordering between $K^{*0}$ and $K^{*\pm}$ is opposite to such
naive expectation. Although the reason behind a difference between
$K^{*0}$ and $K^{*\pm}$ $\rho_{00}$ is not understood yet, but these
species might have different contributions from the vector meson strong
force field. More inputs from theory are required to better understand
the underlying physics mechanisms.

\section{Summary and conclusion}
In summary, the measurements of $\phi$ and $K^{*0}$ $\rho_{00}$ in
Au+Au collisions from RHIC BES-I reveal a surprising pattern with a large positive deviation
from $\frac{1}{3}$ for $\phi$  mesons and no obvious deviation for $K^{*0}$. At present, a fluctuating vector
meson strong force field can accommodate the large positive deviation for
$\phi$ mesons, while more theory inputs are needed for $K^{*0}$. The
recent high statistics RHIC isobar collision (Ru+Ru and Zr+Zr) data
offer a new opportunity to extend the measurement of $\rho_{00}$ for
$K^{*0}$, $\overline{K^{*0}}$, $K^{*+}$, and $K^{*-}$ vector mesons with high precision. We observe the first non-zero
spin alignment for $K^{*\pm}$ in heavy-ion collisions. The $K^{*\pm}$ $\rho_{00}$ is larger
than that of $K^{*0}$ for 20-60\% central isobar collisions. The current large
deviation of $K^{*\pm}$ $\rho_{00}$ and its ordering with
$K^{*0}$ is surprising, and opposite to the naive expectation
from $B$-field. These results pose challenges to current understanding and
inputs from theory are required to interpret the $\rho_{00}$ results
from isobar data.

%\begin{figure}[htb]
%\centerline{%
%\includegraphics[width=12.5cm]{Fig1}}
%\caption{Plot of ...}
%\label{Fig:F2H}
%\end{figure}

%\bibliographystyle{unsrtnat}
%\bibliographystyle{unsrtnat}
%% Inspire style reference is used
%\bibliography{reference.bib}

%\end{linenumbers}
\end{document}